\documentstyle[11pt,aaspp4,epsf]{article}

\newcommand{\etal}{~et~al.}

\newcommand{\wisk}[1]{\ifmmode{#1}\else{$#1$}\fi}

\begin{document}

\vspace{-2.0truecm}
\begin{flushright}
KSUPT-98/2, KUNS-1505 \hspace{0.5truecm} May 1998
\end{flushright}
\vspace{-0.5truecm}

\title{ARGO CMB Anisotropy Measurement Constraints on
  Open and Flat-$\Lambda$ CDM Cosmogonies}

\author{
  Bharat~Ratra\altaffilmark{1}, 
  Ken~Ganga\altaffilmark{2},  
  Rados{\l}aw~Stompor\altaffilmark{3,4},
  Naoshi~Sugiyama\altaffilmark{5},
  Paolo~de~Bernardis\altaffilmark{6},
  and
  Krzysztof~M.~G\'orski\altaffilmark{7,8}
  }

\altaffiltext{1}{Department of Physics, Kansas State University,
                 Manhattan, KS 66506.}
\altaffiltext{2}{IPAC, MS 100--22, California Institute of Technology,
                 Pasadena, CA 91125.}
\altaffiltext{3}{Institute of Astronomy, University of Cambridge,
                 Madingley Road, Cambridge, CB3 0HA, UK.}
\altaffiltext{4}{Copernicus Astronomical Center, Bartycka 18, 
                 00-716 Warszawa, Poland.}
\altaffiltext{5}{Department of Physics, Kyoto University,
                 Kitashirakawa-Oiwakecho, Sakyo-ku, Kyoto 606-8502, Japan.}
\altaffiltext{6}{Departimento di Fisica, Universit\`a di Roma La Sapienza,
                 P. le A. Moro 2, I-00185 Roma, Italy.}
\altaffiltext{7}{Theoretical Astrophysics Center, Juliane Maries Vej 30,
                 2100 Copenhagen \O, Denmark.}
\altaffiltext{8}{Warsaw University Observatory, Aleje Ujazdowskie 4, 
                 00-478 Warszawa, Poland.}

\begin{abstract}
  We use data from the ARGO cosmic microwave background (CMB) anisotropy
  experiment to constrain cosmogonies. We account for the ARGO beamwidth
  and calibration uncertainties, and marginalize over the offset 
  removed from the data. Our derived amplitudes of the CMB 
  anisotropy detected by the ARGO experiment are smaller than those derived 
  previously. 
 
  We consider open and spatially-flat-$\Lambda$ cold dark matter cosmogonies,
  with clustered-mass density parameter $\Omega_0$ in the 
  range 0.1--1, baryonic-mass density parameter $\Omega_B$ in the range
  (0.005--0.029)$h^{-2}$, and age of the universe $t_0$ in the range   
  (10--20) Gyr. Marginalizing over all parameters but $\Omega_0$, the 
  ARGO data favors an open (spatially-flat-$\Lambda$) model with $\Omega_0=$  
  0.23 (0.1). However, these numerical values are model dependent.

  At the 2 $\sigma$ confidence level model normalizations deduced from the
  ARGO data are consistent with those drawn from the UCSB South Pole 1994,
  MAX 4+5, White Dish, and SuZIE data sets. The ARGO open model normalizations
  are also consistent with those deduced from the DMR data. However, for most
  spatially-flat-$\Lambda$ models the DMR normalizations are more than 2 
  $\sigma$ above the ARGO ones.
\end{abstract}

\keywords{cosmic microwave background---cosmology: observations---large-scale
  structure of the universe}

\section{Introduction}

Ganga et al. (1997a, hereafter GRGS) developed a general method to account for 
experimental and observational uncertainties, such as those in the beamwidth
and the calibration, in likelihood analyses of cosmic 
microwave background (CMB) anisotropy data sets. In conjunction with 
theoretically-predicted CMB anisotropy spectra, this method has been used to
account for beamwidth and calibration uncertainties in analyses of the 
Gundersen et al. (1995) UCSB South Pole 1994 data, the Church et al. (1997)
SuZIE data, the MAX 4+5 data (Tanaka et al. 1996; Lim et al. 1996, and 
references therein), and the Tucker et al. (1993) White Dish data (GRGS; Ganga
et al. 1997b, 1998; Ratra et al. 1998).

In this paper we present results from a similar analysis of the ARGO CMB 
anisotropy data set, but now consider a much larger range of cosmological-model
parameter space. The 1993 flight of the balloon-borne ARGO experiment 
resulted in two data sets. One was from data taken in the 
direction of Hercules (de Bernardis et al. 1994a, hereafter deB94), and the 
second consisted of data taken in the direction of Aries and Taurus (Masi et al.
1996). In addition to the CMB signal, the Aries and Taurus data set
is known to have a significant foreground dust signal in the shorter
wavelength data with milder contamination in the longest wavelength
data (Masi et al. 1995, 
1996). To use the Aries and Taurus data to investigate CMB anisotropy one must
carefully model this foreground dust contamination (Masi et al. 1996). In this
paper we therefore only consider the deB94 Hercules data. 
De Bernardis et al. (1993) and Palumbo et al. (1994) describe the ARGO 
telescope and detectors. De Bernardis et al. (1994b) discuss the constraints 
that follow from the ARGO Hercules data on 
some cold dark matter (CDM) and other cosmological models.

ARGO data were taken in four wavelength bands centered at 0.5, 0.8, 1.2,
and 2.0 mm (deB94). While there is dust contamination
in the shorter wavelength data, the longest wavelength data is thought
to be pure CMB anisotropy (deB94). Following deB94 we use only the 2 mm data 
for our CMB anisotropy analyses here.

The FWHM of the beam, assumed to be gaussian, is $52^\prime$, with a $3\%$
one standard deviation uncertainty. While observing, the beam was 
sinusoidally chopped with a half peak-to-peak chop amplitude of $0.9^\circ$
on the sky. 63 points were observed with the payload 
performing azimuth step and integrate scans at constant elevation. This 
scan strategy, combined with sky rotation, produced a W-like scan pattern in
equatorial coordinates (Figure 1 of deB94), sparsely sampling a region 1 hour 
wide in right 
ascension and 15$^\circ$ wide in declination. deB94 
remove a single offset from the entire data set prior to binning. The ARGO
Hercules scan absolute calibration uncertainty is $5\%$ (one standard
deviation)\footnote{
The value quoted in deB94, $15\%$, is a higher standard deviation calibration
uncertainty.}. 

ARGO data is able to robustly constrain the normalization of the CMB
spectra. However, ARGO alone is unable to strongly constrain the other
cosmological parameters considered. Robust constraints on these 
cosmological parameters will require a combined likelihood analysis 
of many CMB anisotropy data sets (including ARGO). In this paper we use
the ARGO data set as a basis for developing a formalism that will
eventually allow such a combined analysis. 

In $\S$2 we summarize the computational techniques used in our 
analysis. See GRGS for a more detailed description. In $\S$2 we also
describe the greatly extended cosmological parameter space we consider here.
Results are given and discussed in $\S$3. Conclusions are presented in $\S$4.

\section{Summary of Computation and Models Considered}

The reduced 2 mm Hercules data is shown in Figure 1. The zero-lag
window function for these observations is shown in Figure 2 and the 
zero-lag window function parameters are listed in Table 1.

Figure 2 also shows some of the model CMB anisotropy spectra used 
in the analysis. In earlier analyses (GRGS; Ganga et al. 1997b, 1998;
Ratra et al. 1998), only a few (25) spectra were
used. In this paper we use 798 spectra to cover a greatly extended range of
the cosmological parameter space of the open
and spatially-flat-$\Lambda$ CDM models, and cover it with 
a higher resolution grid than previously used. This allows for 
an explicit construction of the full likelihood function as a function
of all the cosmological parameters considered, and for making the various 
likelihood functions derived by marginalizing over parameters. This 
had not previously been possible. 

We focus here on a spatially open CDM model and
a spatially flat CDM model with a cosmological constant $\Lambda$. These
low density models are consistent with most current observational 
constraints. For discussions see Coles et al. (1998), Bartelmann et al. (1998),
Jenkins et al. (1998), Park et al. (1998), Merch\'an et al. (1998),
Cole et al. (1998), Cavaliere, Menci, \& Tozzi (1998), and Somerville 
\& Primack (1998). 

The models have gaussian, adiabatic primordial energy-density power
spectra. The flat-$\Lambda$ model CMB anisotropy computations use a 
scale-invariant energy-density perturbation power spectrum (Harrison 1970;
Peebles \& Yu 1970; Zel'dovich 1972), as predicted in the simplest 
spatially-flat inflation models (Guth 1981; Kazanas 1980; Sato 1981a,b).
The open model computations use the energy-density power spectrum 
(Ratra \& Peebles 1994, 1995; Bucher, Goldhaber, \& Turok 1995; Yamamoto,
Sasaki, \& Tanaka 1995) predicted in the simplest open-bubble 
inflation models (Gott 1982; Guth \& Weinberg 1983). The computation of the 
CMB anisotropy spectra is described in Stompor (1994) and Sugiyama (1995). 

The spectra are parameterized by their quadrupole-moment
amplitude $Q_{\rm rms-PS}$, the clustered-mass density parameter $\Omega_0$,
the baryonic-mass density parameter $\Omega_B$, and the age of the 
universe $t_0$. We have evaluated the spectra for a range of
$\Omega_0$ spanning the interval 0.1 to 1 in steps of 0.1, for a range
of $\Omega_B h^2$ [the Hubble parameter
$h = H_0/(100 h\ {\rm km}\ {\rm s}^{-1}\ {\rm Mpc}^{-1})$] spanning the 
interval 0.005 to 0.029 in steps of 0.004, and for a range of $t_0$ spanning 
the interval 10 to 20 Gyr in steps of 2 Gyr. Current observational estimates
are more consistent with $\Omega_0 \sim 0.4$ (e.g., Cole et al. 1997; Eke
et al. 1998), with $t_0 \sim 12$ Gyr (e.g., Reid 1997; Gratton et al. 1997),
and disfavor $\Omega_B h^2$ larger than $\sim 0.02$ (e.g., Fukugita, Hogan,
\& Peebles 1998; Burles \& Tytler 1998).

While it is of interest to also consider other cosmological parameters,
such as tilt or gravity wave fraction, to make the problem tractable we
have focussed on the four parameters mentioned above. We emphasize however that 
the results of the analysis are model dependent. For instance, tilted 
flat-$\Lambda$ models at fixed $\Omega_B h^2$ would likely lead to a different 
constraint on $\Omega_0$ than that derived below in the scale-invariant
flat-$\Lambda$ model with varying $\Omega_B h^2$.

GRGS describe the computation of the likelihood function. Since
deB94 remove an offset from the data, we assume a
uniform prior in the amplitude of the offset removed
and marginalize over this amplitude when computing the likelihood function 
(Bond et al. 1991, Bunn et al. 1994; GRGS; Church et al. 1997; Ratra et al. 
1998). This must be done since the removal
of the offset also removes an undetermined
amount of the CMB anisotropy signal from the data. Beamwidth and calibration 
uncertainties are accounted for as described in GRGS. 

The open and flat-$\Lambda$ model likelihoods are a function of
four parameters: $Q_{\rm rms-PS}$, $\Omega_0$, $\Omega_B h^2$, and $t_0$. 
We also compute marginalized likelihood functions by integrating over one or
more of these parameters after assuming a uniform prior in the relevant parameters.
The prior is set to zero outside the ranges considered for the parameters.

To derive central values and limits from the likelihood functions we 
assume a uniform prior in the relevant parameter, so the corresponding 
posterior probability density distribution function vanishes outside the
chosen parameter range and is equal to the likelihood function inside
this range. The deduced central value of the parameter is taken to be the 
value at which the posterior probability density peaks. The limits we 
quote are based on the highest posterior density (HPD) prescription. They 
are determined by
integrating the posterior probability density starting at the peak and 
minimizing the difference between the upper and lower limits. The $\pm 1$
$\sigma$ and $\pm 2$ $\sigma$ HPD limits encompass 68.3\% and 95.5\%
of the area, respectively. See GRGS for further details. Of course, the 
quoted limits depend on the prior range considered for the parameter, if the 
likelihood function is not sharply peaked within the range considered.
This is the case for a number of the likelihood functions obtained below.

\section{Results and Discussion}

For the flat bandpower spectrum the ARGO likelihood
peaks at bandtemperature $\delta T_l = 33\ \mu$K, with a 1 $\sigma$ range
of $28\ \mu{\rm K} < \delta T_l < 38\ \mu{\rm K}$ and likelihood ratio
= $ 1 \times 10^{21}$. For the fiducial CDM model spectrum  
$\delta T_l = 30\ \mu$K, with 1 $\sigma$ range $26\ \mu{\rm K} < \delta T_l
< 35\ \mu{\rm K}$ and likelihood ratio $5 \times 10^{21}$. These numerical 
values account for the marginalization over the offset
removed from the data, as well as the beamwidth and calibration uncertainties.
For fiducial CDM deB94 find that the likelihood peaks at $\delta T_l 
= 39\ \mu$K, with 
1 $\sigma$ range $34\ \mu{\rm K} < \delta T_l < 45\ \mu{\rm K}$ and 
likelihood ratio = $ 5 \times 10^{23}$. The deB94 1 $\sigma$ range accounts for
calibration uncertainty by adding it in quadrature to the error bars derived
ignoring it, but ignores offset removal and beamwidth uncertainty. The 
difference between the numerical values derived in deB94 and
here is mostly because they have not marginalized over the amplitude of
the offset removed while we have. 

For the flat bandpower spectrum the ARGO average 
1 $\sigma$ $\delta T_l$ error bar is $\sim 15\%$: ARGO data results in a very 
significant detection of CMB anisotropy, even after accounting for beamwidth 
and calibration uncertainties. For comparison, the corresponding DMR 
error bar is $\sim 10-12\%$ (depending on model, G\'orski et al. 1998). 
The MAX 4+5 data set also results in a small error bar $\sim 14\%$
(Ganga et al. 1998). 

We note that $\delta T_l$ values estimated using the flat bandpower and 
fiducial CDM spectra are not identical.
The variation in the deduced $\delta T_l$ values from model to model is an 
indication of the accuracy of the flat bandpower approximation to the real
spectrum over the range of the window function of the 
experiment. The variation from model to model found here is comparable to 
that found from the SP94 and MAX 4+5 data (GRGS; Ganga et al. 1998) but is
smaller than that found from the SuZIE and White Dish data (Ganga et al.
1997b; Ratra et al. 1998).

For both the open and flat-$\Lambda$ models, the four-dimensional posterior
probability density distribution function $L(Q_{\rm rms-PS}, \Omega_0,
\Omega_B h^2, t_0)$ is nicely peaked in the $Q_{\rm rms-PS}$ direction
but quite flat in the other three directions. The dotted lines in Figure
3 illustrate this flatness in the $(\Omega_0, \ t_0)$ subspace of the 
flat-$\Lambda$ model. Note that it is possible to distinguish between regions
of parameter space at only slightly better than 0.25 $\sigma$ confidence. 
The irregular 
solid lines in Figure 5 are the 2 $\sigma$ contours of the four-dimensional 
posterior distribution projected on to the $(\Omega_0, \ Q_{\rm rms-PS})$
subspace. They clearly show the steepness and peak (solid circles) in the 
$Q_{\rm rms-PS}$ direction.

Marginalizing over $Q_{\rm rms-PS}$ results in a three-dimensional posterior
distribution $L(\Omega_0,  \Omega_B h^2, t_0)$ which is steeper. The dashed
lines in Figure 3 are the contours of this function. Note that it is now 
possible to distinguish between regions of parameter space at better than 
1 $\sigma$ confidence. Marginalizing over one more parameter (in addition 
to $Q_{\rm rms-PS}$) results in two-dimensional posterior probability
distribution functions which are peaked (albeit, in most cases, at an
edge of the parameter range considered, see the solid circle in Figure 3).
It is now possible to distinguish between parts of parameter space at better
than 3 $\sigma$  confidence --- see the solid lines in Figure 3.

As discussed below, caution must be exercised when interpreting the 
discriminative power of these formal limits, since they depend 
sensitively on the values of the parameters beyond which the uniform prior 
has been set to zero. 

Figure 3 also illustrates a point noted earlier in the analysis of
the MAX 4+5 data set (Ganga et al. 1998): conclusions about the most
favored model drawn from the full four-dimensional posterior distribution
tend to differ from those deduced from the three-dimensional posterior
distribution derived by marginalizing the four-dimensional one over 
$Q_{\rm rms-PS}$. For example, from a similar plot of the $(\Omega_0, \
\Omega_B h^2)$  subspace of the open model (not shown), one finds that the 
four-dimensional distribution favors a large value of $\Omega_0 \sim 1$,
while the three-dimensional distribution favors a small $\Omega_0 \sim 0.25$.
As discussed in Ganga et al. (1998), this is a consequence of the 
asymmetry of the posterior distribution. 
This is not a significant issue since at the 2 $\sigma$ 
confidence level the four-dimensional posterior distribution is flat in the
$\Omega_0$ direction and so it is not statistically meaningful to discuss
how it varies with $\Omega_0$.

As mentioned above, the two-dimensional posterior 
distributions (derived by marginalizing the four-dimensional distribution
over two parameters at a time) allows one to distinguish between regions of 
parameter space at a higher level of confidence. Figure 4 illustrates this 
for the three 
cosmological parameters ($\Omega_0$, $\Omega_B h^2$, $t_0$), for both the 
open and flat-$\Lambda$ models. Certain parts of parameter space can now 
be formally ruled out at better than 2 $\sigma$ significance. For example, for 
the open
model a region in parameter space centered near $\Omega_0 \sim 0.7$, 
$\Omega_B h^2 \sim 0.005$, and $t_0 \sim 10$ Gyr is ruled out at 3 $\sigma$.
Again, caution must be exercised when interpreting the discriminative 
power of such formal limits. 

Figure 5 shows the contours of the two-dimensional posterior distribution 
for $Q_{\rm rms-PS}$ and $\Omega_0$, derived by marginalizing the 
four-dimensional distribution over $\Omega_B h^2$ and $t_0$. These are shown
for the ARGO, DMR, SP94, MAX 4+5, White Dish, and SuZIE data sets, for 
both the open and flat-$\Lambda$ models. For all but the DMR data we also show 
2 $\sigma$ confidence contours determined from projecting the four-dimensional
posterior distribution on to this ($Q_{\rm rms-PS}$, $\Omega_0$) subspace.
Clearly, at 2 $\sigma$, constraints on these parameters derived from the
ARGO data are mostly consistent with those derived from the other data sets.
However, the DMR and ARGO data are more consistent for an open model than for 
the flat-$\Lambda$ case, panels $a)$ and $b)$ of Figure 5. In fact, there
is very little overlap between the 2 $\sigma$ ranges (derived from the 
two-dimensional posterior distributions) of the ARGO and DMR normalizations
for the flat-$\Lambda$ models. There is, however, significant overlap between
the 2 $\sigma$ ranges derived by projecting the ARGO and DMR four-dimensional
posterior distributions\footnote{
Since the DMR four-dimensional posterior distribution is only very weakly 
dependent on $\Omega_B h^2$ and $t_0$, the confidence contours derived 
from this distribution projected on to the ($Q_{\rm rms-PS}$, $\Omega_0$)
plane are close to those derived from this distribution marginalized 
over $\Omega_B h^2$ and $t_0$.}. The SP94 results shown in panels $c)$
and $d)$ are those derived from the full Ka+Q data set. Sample variance
and noise considerations indicate that the SP94 Ka band data is more 
consistent with what is expected for CMB anisotropy data (GRGS). At fixed
$\Omega_0$ the Ka band data results in a lower deduced $Q_{\rm rms-PS}$
(compared to that from the Ka+Q data, GRGS), so the SP94 Ka band results
are more consistent with the ARGO results. The MAX 4+5 results shown in 
panels $e)$ and $f)$ of Figure 5 are those derived from the MAX 4 ID and
SH and MAX 5 HR, MP, and PH data sets. Sample variance and noise 
considerations indicate that the MAX 4 ID and MAX 5 HR data are more consistent
with what is expected for CMB anisotropy data (Ganga et al. 1998). This subset
of the MAX 4+5 data has a lower $Q_{\rm rms-PS}$ (Ganga et al. 1998) and is 
thus more consistent with the ARGO data. 

The Python (Platt et al. 1997) and Saskatoon (Netterfield et al. 1997)
experiments are also sensitive to angular scales probed by ARGO.
The Python large-chop result and some of the relevant Saskatoon $n$-point
chop results are larger than the ARGO result. On the other hand, some
of the Saskatoon $n$-point chop results are consistent with the ARGO 
result. 

Figure 6 shows the one-dimensional posterior distribution
functions for $\Omega_0$, $\Omega_B h^2$, $t_0$, and $Q_{\rm rms-PS}$, 
derived by marginalizing the four-dimensional posterior distribution over the
other three parameters. From these one-dimensional posterior distributions,
ARGO data favors an open (flat-$\Lambda$) model with either $\Omega_0$ = 0.23 
(0.10), or
$\Omega_B h^2$ = 0.029 (0.020), or $t_0$ = 20 (12) Gyr, amongst the models
considered. We emphasize that each of these are derived from one-dimensional
posterior distributions and thus can not be simultaneously imposed.
Also shown in Figure 6 are the limits derived from the
one-dimensional distributions and the (projected) four-dimensional distribution.
At 2 $\sigma$ confidence the ARGO data only formally rules out small regions of 
parameter space. Specifically, from the one-dimensional posterior 
distributions, the ARGO data requires $\Omega_0$ $< 0.69$ or $> 0.76$
($\Omega_0$ $< 0.91$), or $\Omega_B h^2$ $> 0.006$ ($\Omega_B h^2$ $> 0.006$),
or $t_0$ $> 10$ Gyr ($t_0$ $< 20$ Gyr) for the open (flat-$\Lambda$) model at
2 $\sigma$. Since some of the one-dimensional posterior distributions 
peak at the edge of the parameter range considered some of these limits
must be considered to be formal. Less controversially, it is clear that, 
for both open and flat-$\Lambda$ models, ARGO data favors a low-density 
universe, although $\Omega_0 = 1$ is not strongly ruled out. Similarly, 
ARGO data also mildly favors a relatively large $\Omega_B h^2$ or a young (old)
flat-$\Lambda$ (open) model. As discussed above, ARGO results in fairly
tight constraints on $Q_{\rm rms-PS}$ (panels $g)$ and $h)$ of Figure 6),
and these are more consistent with the DMR results for the open model
than for the flat-$\Lambda$ case.

Care is needed when interpreting the discriminative 
power of such formal limits. Consider a posterior density function which is
a gaussian and nicely peaked inside the parameter range considered. The 1 
and 2 $\sigma$ HPD limits for such a gaussian correspond to a value of
the posterior distribution relative to that at the peak of 0.61 and 0.14
respectively. For the open model posterior distributions shown in
Figure 6, the 1 and 2 $\sigma$ HPD limits correspond to values of the 
posterior distribution relative to that at the peak of 0.63 \& 0.56
($\Omega_0$), 0.95 \& 0.91 ($\Omega_B h^2$), 0.93 \& 0.86 ($t_0$), 
and 0.36 \& 0.13 ($Q_{\rm rms-PS}$). It is hence probably fairer to 
conclude that the formal ARGO statistical limits on $\Omega_B h^2$ and
$t_0$, and the 2 $\sigma$ limits on $\Omega_0$, should be taken much less 
seriously than those on $Q_{\rm rms-PS}$, and the 1 $\sigma$ limits on
$\Omega_0$.

\section{Conclusion}

In our likelihood analyses of the ARGO Hercules scan CMB anisotropy
data we have explicitly accounted for beamwidth and calibration uncertainties
and have marginalized over the amplitude of the offset
removed from the data. As a consequence the results derived here differ from 
those derived earlier. The ARGO results are mostly consistent with those 
derived from the DMR, SP94, MAX 4+5, White Dish and SuZIE data sets. 
We emphasize again that the ARGO results, as well as those derived for the 
other data sets, depend on the models considered.

While the ARGO data does significantly constrain $Q_{\rm rms-PS}$ (and mildly 
constrain $\Omega_0$), robust constraints on these and other cosmological
parameters from the CMB anisotropy must await a models-based combined 
likelihood analysis of many different data sets.

\bigskip

We acknowledge helpful discussions with S. Masi, A. Noriega-Crespo,  
and G. Rocha. This work was partially
carried out at the Infrared Processing and Analysis Center and the Jet
Propulsion Laboratory of the California Institute of Technology, under a
contract with the National Aeronautics and Space Administration.
BR acknowledges support from NSF grant EPS-9550487 with matching support from 
the state of Kansas and from a K$^*$STAR First award. RS acknowledges 
support from a UK PPARC grant and from Polish Scientific Committee (KBN)
grant 2P03D00813.


\begin{table}
\begin{center}
\caption{Numerical Values for the Zero-Lag Window Function 
Parameters\tablenotemark{a}}
\vspace{0.3truecm}
\tablenotetext{{\rm a}}{The value of $l$ where $W_l$ is
largest, $l_{\rm m}$, the two values of $l$ where $W_{l_{e^{-0.5}}} =
e^{-0.5} W_{l_{\rm m}}$, $l_{e^{-0.5}}$, the effective multipole,
$l_{\rm e} = I(lW_l)/I(W_l)$, and 
$I(W_l) = \sum^\infty_{l=2}(l+0.5)W_l/\{l(l+1)\}$.}
\begin{tabular}{lccccc}
\tableline\tableline
$l_{e^{-0.5}}$ & $l_{\rm e}$ & $l_{\rm m}$ &
  $l_{e^{-0.5}}$ & $\sqrt{I(W_l)}$  \\
\tableline
60 & 97.6 & 109 & 168 & 0.551 \\
\tableline
\end{tabular}
\end{center}
\end{table}


\clearpage
\centerline{\bf Figure Captions}

\medskip
\noindent
Fig.~1.--
{\protect Measured 2 mm thermodynamic temperature differences (with $\pm 
    1$-$\sigma$ error bars) on the sky as a function of scan position
    in the direction of Hercules. Note that the 63 points in the scan are
    at varying right ascension and declination.
}

\medskip
\noindent
Fig.~2.--
{\protect CMB anisotropy multipole moments $l(l+1)C_l/(2\pi )\times
    10^{10}$ (solid lines, scale on left axis, note that these are fractional 
    anisotropy moments and thus dimensionless) as a function of
    multipole $l$, for selected models normalized to the DMR maps
    (G\'orski et al. 1998; Stompor 1997). Panels $a)-c)$ show selected
    flat-$\Lambda$ models. The heavy lines are the $\Omega_0 = 0.1$,
    $\Omega_B h^2 = 0.021$, and $t_0 = 12$ Gyr case, which is close to 
    where the likelihoods (marginalized over all but one parameter at
    a time) are at a maximum. Panel $a)$ shows five $\Omega_B h^2$ = 0.021,
    $t_0$ = 12 Gyr models with $\Omega_0$ = 0.1, 0.3, 0.5, 0.7, and 0.9
    in descending order at the $l \sim 200$ peaks. Panel $b)$ shows seven 
    $\Omega_0$ = 0.1, $t_0$ = 12 Gyr models with $\Omega_B h^2$ = 0.029,
    0.025, 0.021, 0.017, 0.013, 0.009, and 0.005 in descending order at
    the $l \sim 200$ peaks. Panel $c)$ shows six $\Omega_0$ = 0.1, 
    $\Omega_B h^2$ = 0.021 models with $t_0$ = 20, 18, 16, 14, 12, and
    10 Gyr in descending order at the $l \sim 200$ peaks. Panels $d)-f)$ 
    show selected open models. The heavy lines are the $\Omega_0 = 0.2$,
    $\Omega_B h^2 = 0.029$, and $t_0 = 20$ Gyr case, which is close to 
    where the likelihoods (marginalized over all but one parameter at
    a time) are at a maximum. Panel $d)$ shows five $\Omega_B h^2$ = 0.029,
    $t_0$ = 20 Gyr models with $\Omega_0$ = 1, 0.8, 0.6, 0.4, and 0.2
    from left to right at the peaks (the peak of the $\Omega_0$ = 0.2
    model is off scale). Panel $e)$ shows seven 
    $\Omega_0$ = 0.2, $t_0$ = 20 Gyr models with $\Omega_B h^2$ = 0.029,
    0.025, 0.021, 0.017, 0.013, 0.009, and 0.005 in descending order at
    $l \sim 600$. Panel $f)$ shows six $\Omega_0$ = 0.2, 
    $\Omega_B h^2$ = 0.029 models with $t_0$ = 20, 18, 16, 14, 12, and
    10 Gyr in descending order at $l \sim 600$.
    Also shown is the ARGO 2 mm zero-lag
    window function $W_l$ (dotted lines, scale on right axis). See Table 1
    for $W_l$-parameter values. The ARGO data mainly constrains the area
    under the product of the model spectrum and the window function on this 
    plot. This is clearly larger for the DMR-normalized flat-$\Lambda$ models
    than for the DMR-normalized open models. ARGO thus favors a lower 
    relative normalization between the flat-$\Lambda$ and open model than
    does the DMR.
}

\medskip
\noindent
Fig.~3.--
{\protect Maxima and confidence contours of various posterior probability
    density distribution functions for the
    $(\Omega_0, t_0)$ subspace of the flat-$\Lambda$ model. Dotted lines
    (solid squares) show the contours (maxima) of the four-dimensional
    $(Q_{\rm rms-PS}, \Omega_0, \Omega_B h^2, t_0)$ posterior distribution;
    contours of 0.1 and 0.25 $\sigma$ confidence are shown. Dashed lines
    (solid triangles) show the contours (maxima) of the three-dimensional
    $(\Omega_0, \Omega_B h^2, t_0)$ posterior distribution (derived by 
    marginalizing the four-dimensional one over $Q_{\rm rms-PS}$). Contours 
    of 0.1, 0.25, 0.5, and 1 $\sigma$ confidence are shown. Solid lines
    (solid circles) show the contours (maxima) of the 
    two-dimensional posterior distribution (derived by 
    marginalizing the four-dimensional one over the other two parameters). 
    Contours of 1, 2, and 3 $\sigma$ confidence are shown (the 3 $\sigma$ 
    contour is not labelled).
}

\medskip
\noindent
Fig.~4.--
   {\protect Confidence contours and maxima of the two-dimensional posterior 
   probability
   density distribution functions, as a function of the two parameters on
   the axes of each panel (derived by marginalizing the four-dimensional 
   posterior distribution over the other two parameters). Dashed lines
   (open circles) show the contours (maxima) of the open case and solid 
   lines (solid circles) show those of the flat-$\Lambda$ model. 
   Contours of 0.25, 0.5, 1, 2, and 3 $\sigma$ confidence are shown
   (3 $\sigma$ contours are not labelled).
}

\medskip
\noindent
Fig.~5.--
{\protect Confidence contours and maxima of the two-dimensional 
   $(Q_{\rm rms-PS}, \Omega_0)$ posterior probability density distribution 
   function. Panels $a)$, $c)$, $e)$, and $g)$ are for the flat-$\Lambda$ 
   model and panels $b)$, $d)$, $f)$, and $h)$ are for the open model. 
   Note the different scale on the vertical $(Q_{\rm rms-PS})$ axis in 
   each pair of panels. Shaded regions show the ARGO results, with denser
   shading for the 1 $\sigma$ confidence region and less-dense shading 
   for the 2 $\sigma$ region. Irregular solid lines show the 2 $\sigma$ 
   confidence contours derived by projecting the four-dimensional ARGO
   posterior distribution in to this plane. [I.e., for each set of values of
   $(Q_{\rm rms-PS}, \Omega_0)$ we check if there is any choice of 
   $\Omega_B h^2$ and $t_0$ such that the point $(Q_{\rm rms-PS}, \Omega_0,
   \Omega_B h^2, t_0)$ is within the 2 $\sigma$ region of the four-dimensional
   ARGO posterior distribution. The projected 2 $\sigma$ limits enclose those
   values of $(Q_{\rm rms-PS}, \Omega_0)$ for which such a point exists.]
   In panels $a)-f)$ hatched areas
   show the two-dimensional posterior probability density distribution 
   function confidence regions for the DMR data (panels $a)$ and $b)$, 
   G\'orski et al. 1998; Stompor 1997), the SP94 Ka+Q data (panels $c)$
   and $d)$, GRGS), and the MAX 4+5 data (panels $e)$ and $f)$, Ganga et al.
   1998). Heavy dashed confidence contours bounding these regions are labelled 
   (except for the DMR cases); denser hatching corresponds to the 1 $\sigma$
   confidence region and less-dense hatching to the 2 $\sigma$ confidence 
   region. In panels $c)-f)$ unlabeled irregular light dashed lines show the 
   2 $\sigma$ confidence contours derived by projecting the SP94 and MAX 4+5
   four-dimensional posterior distributions in to this plane. Panels $g)$ 
   and $h)$ show the SuZIE 2 $\sigma$ upper limit
   (the hatched region bounded by the labelled heavy dashed line, Ganga et 
   al. 1997b) and the White Dish 2 $\sigma$ upper limit (labelled heavy
   dotted line in panel $h)$ and not shown in panel $g)$ since it is 
   off scale, Ratra et al. 1998). The unlabeled irregular light dashed and
   dotted lines in panel $h)$ are the corresponding SuZIE and White Dish 
   2 $\sigma$ confidence upper limits derived by projecting the 
   four-dimensional posterior distribution in to this plane (these limits
   are not shown in panel $g)$ since they are off scale). Solid circles
   show the maxima of the ARGO two-dimensional posterior distribution and
   open circles show those of the other data sets (not shown for DMR, SuZIE,
   and White Dish). The DMR results are a composite of those from 
   analyses of the two extreme data sets: i) galactic frame with quadrupole
   included and correcting for faint high-latitude galactic emission; and
   ii) ecliptic frame with quadrupole excluded and no other galactic 
   emission correction (G\'orski et al. 1998).
}

\medskip
\noindent
Fig.~6.--
{\protect One-dimensional posterior probability density distribution 
   functions for $\Omega_0$, $\Omega_B h^2$, $t_0$, and $Q_{\rm rms-PS}$ 
   (derived by marginalizing the four-dimensional one over the other
   three parameters) in the open and flat-$\Lambda$ models.  These have
   been renormalized to unity at the peaks. Dotted vertical
   lines show the confidence limits derived from these one-dimensional 
   posterior distributions and solid vertical lines in panels $g)$ and
   $h)$ show the $\pm 1$ and $\pm 2$ $\sigma$ confidence limits derived by 
   projecting the four-dimensional ARGO posterior distribution. Note
   that, as discussed at the end of $\S$3, some of these formal limits
   have very little discriminative power. The 2 $\sigma$ DMR 
   (marginalized and projected) confidence limits in panels $g)$ and $h)$ 
   are a composite of those from the 
   two extreme DMR data sets (see caption of Figure 5).
}

\clearpage
\pagestyle{empty}

\begin{figure}[t]
\centerline{\epsfxsize=5.5truein \epsffile{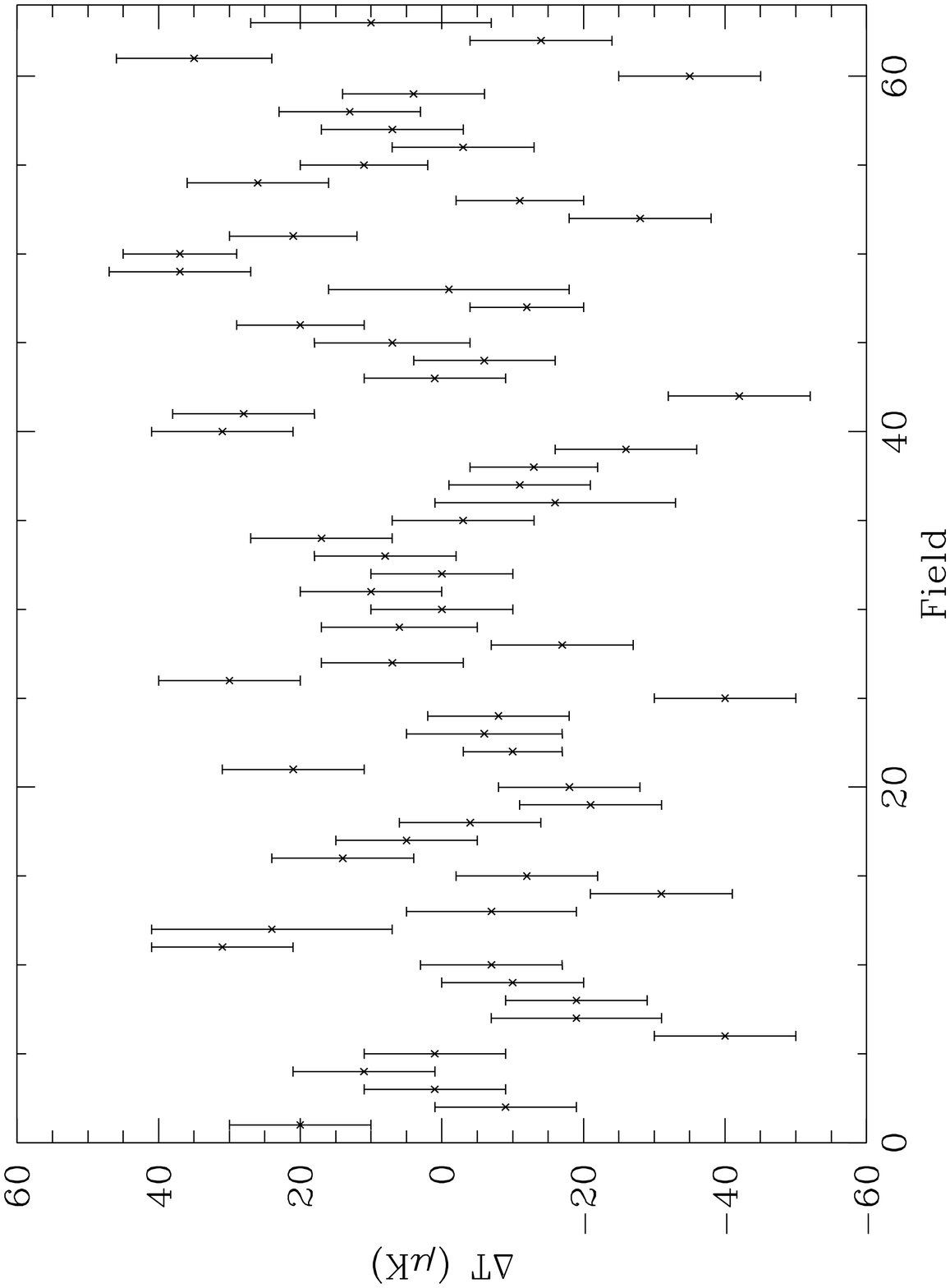}}
\mbox{\large Figure 1}
\end{figure}
\clearpage
\begin{figure}[t]
\centerline{\epsfxsize=5.5truein \epsffile{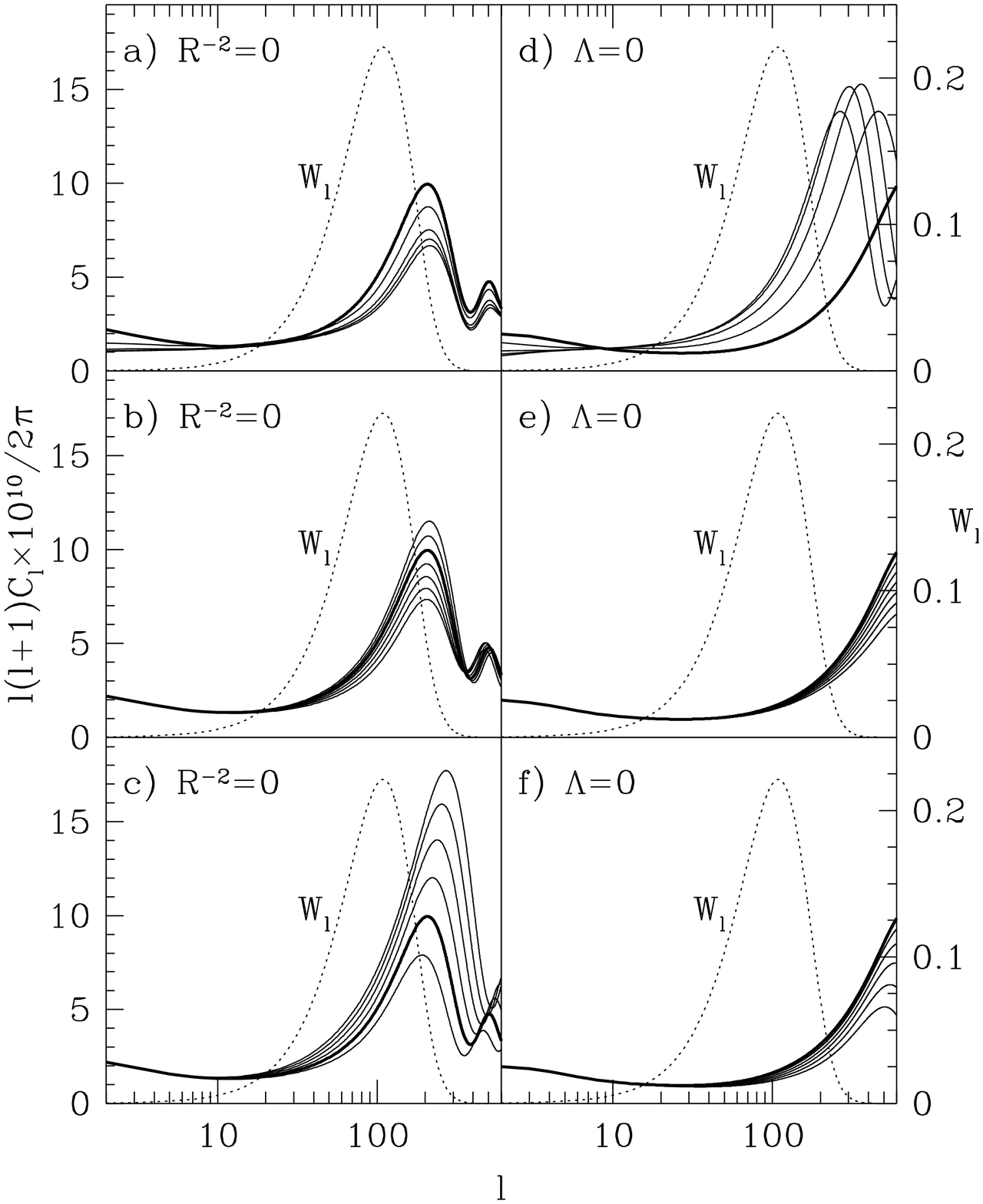}}
\mbox{\large Figure 2}
\end{figure}
\clearpage
\begin{figure}[t]
\centerline{\epsfysize=8.0truein \epsffile{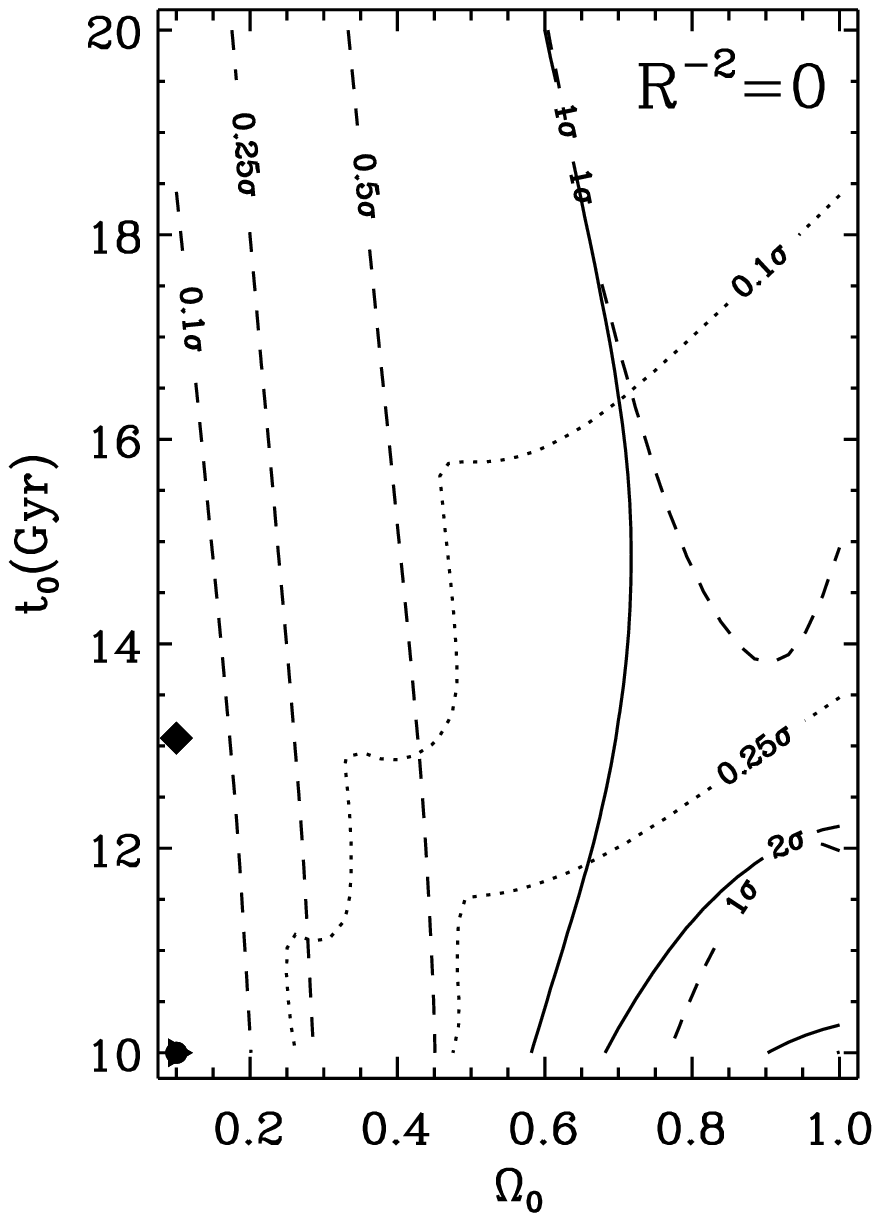}}
\mbox{\large Figure 3}
\end{figure}
\clearpage
\begin{figure}[t]
\centerline{\epsfysize=8.0truein \epsffile{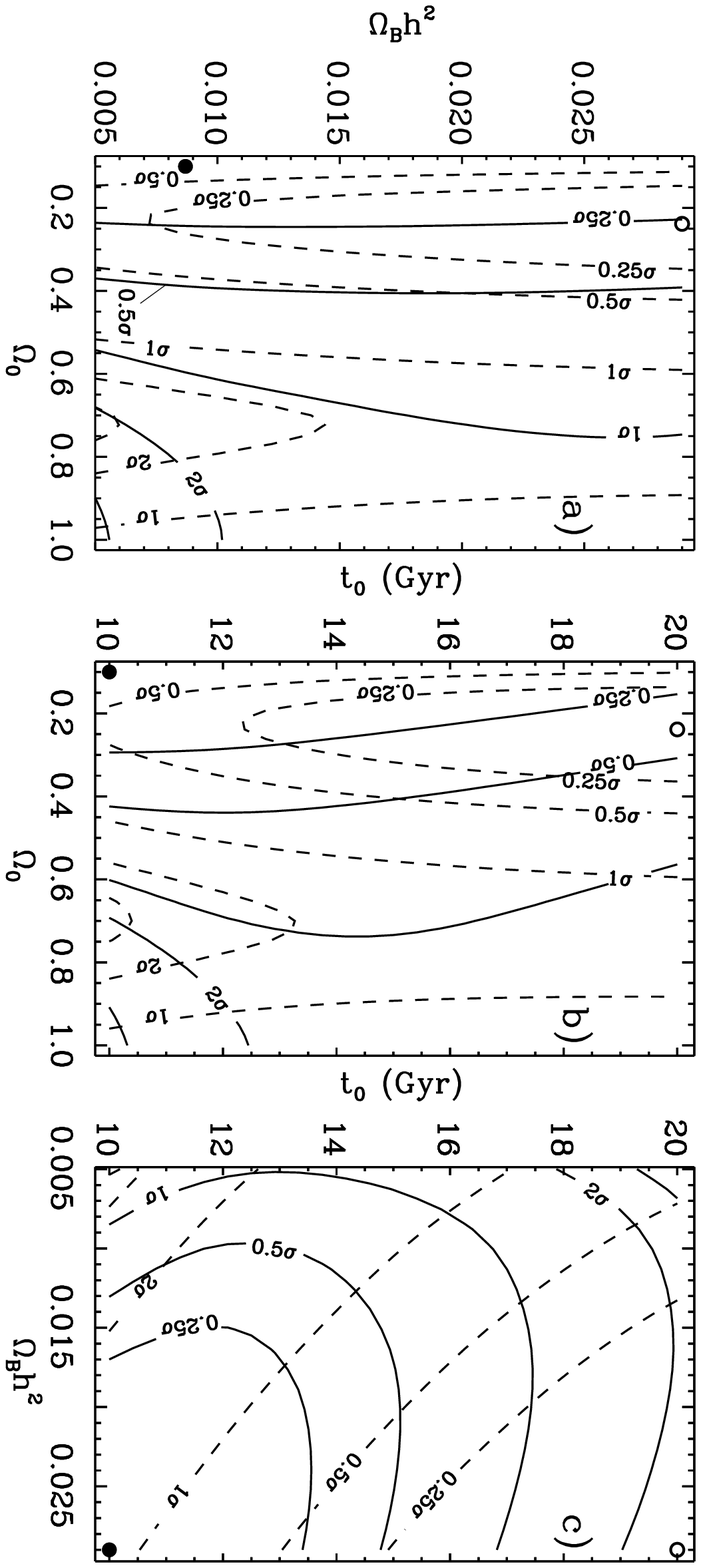}}
\mbox{\large Figure 4}
\end{figure}
\clearpage
\begin{figure}[t]
\centerline{\epsfysize=8.0truein \epsffile{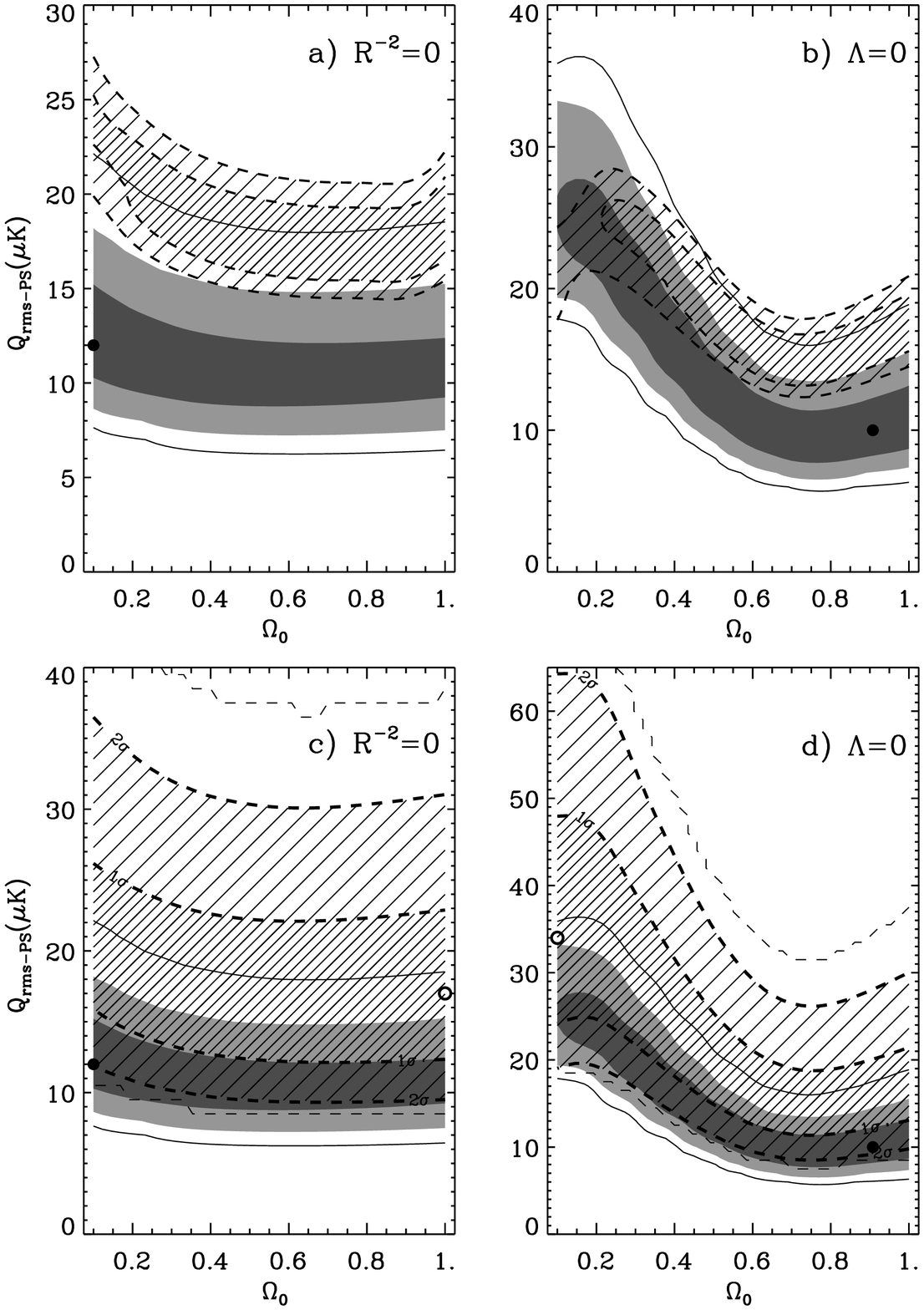}}
\mbox{\large Figure 5.1}
\end{figure}
\clearpage
\begin{figure}[t]
\centerline{\epsfysize=8.0truein \epsffile{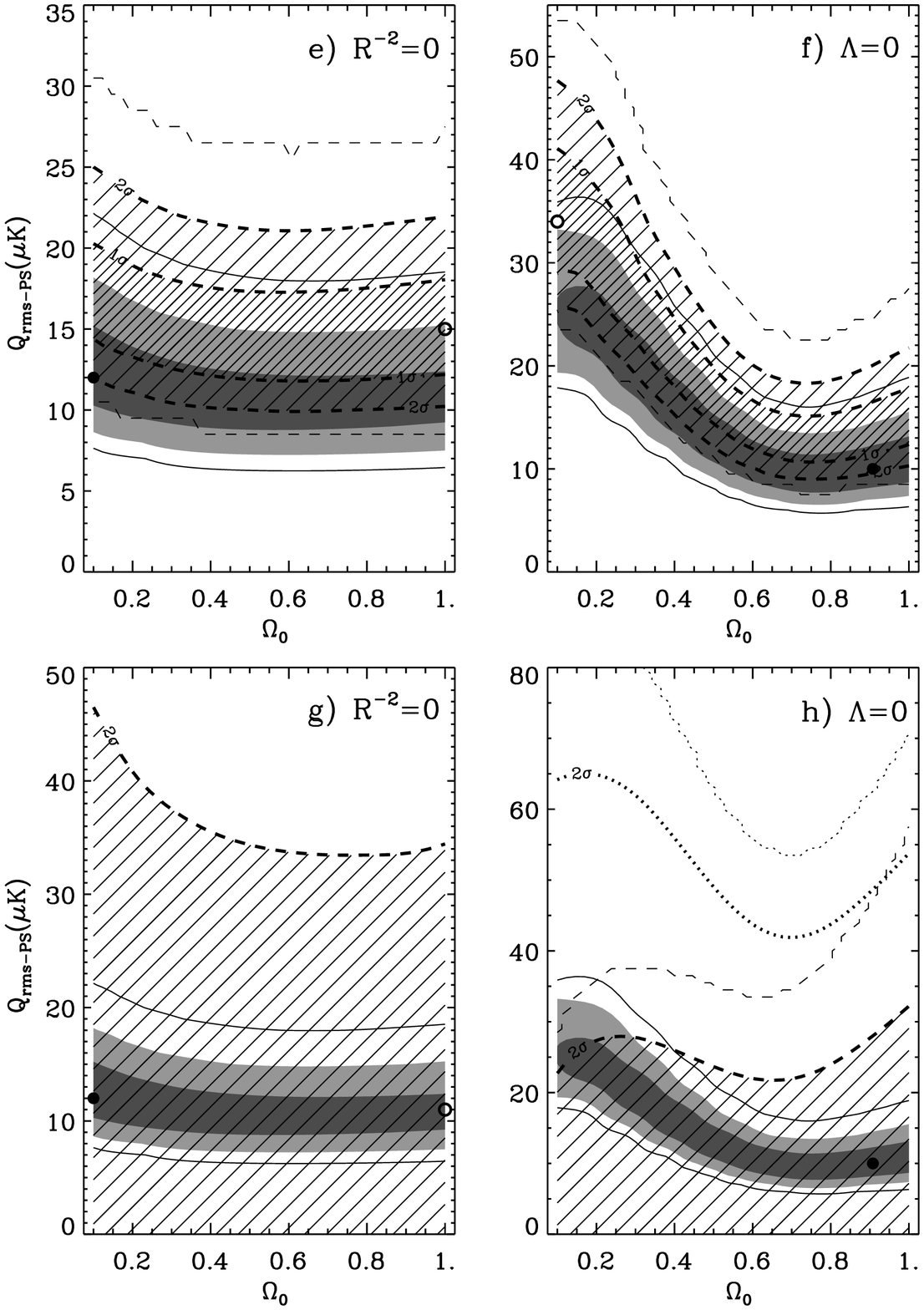}}
\mbox{\large Figure 5.2}
\end{figure}
\clearpage
\begin{figure}[t]
\centerline{\epsfysize=8.0truein \epsffile{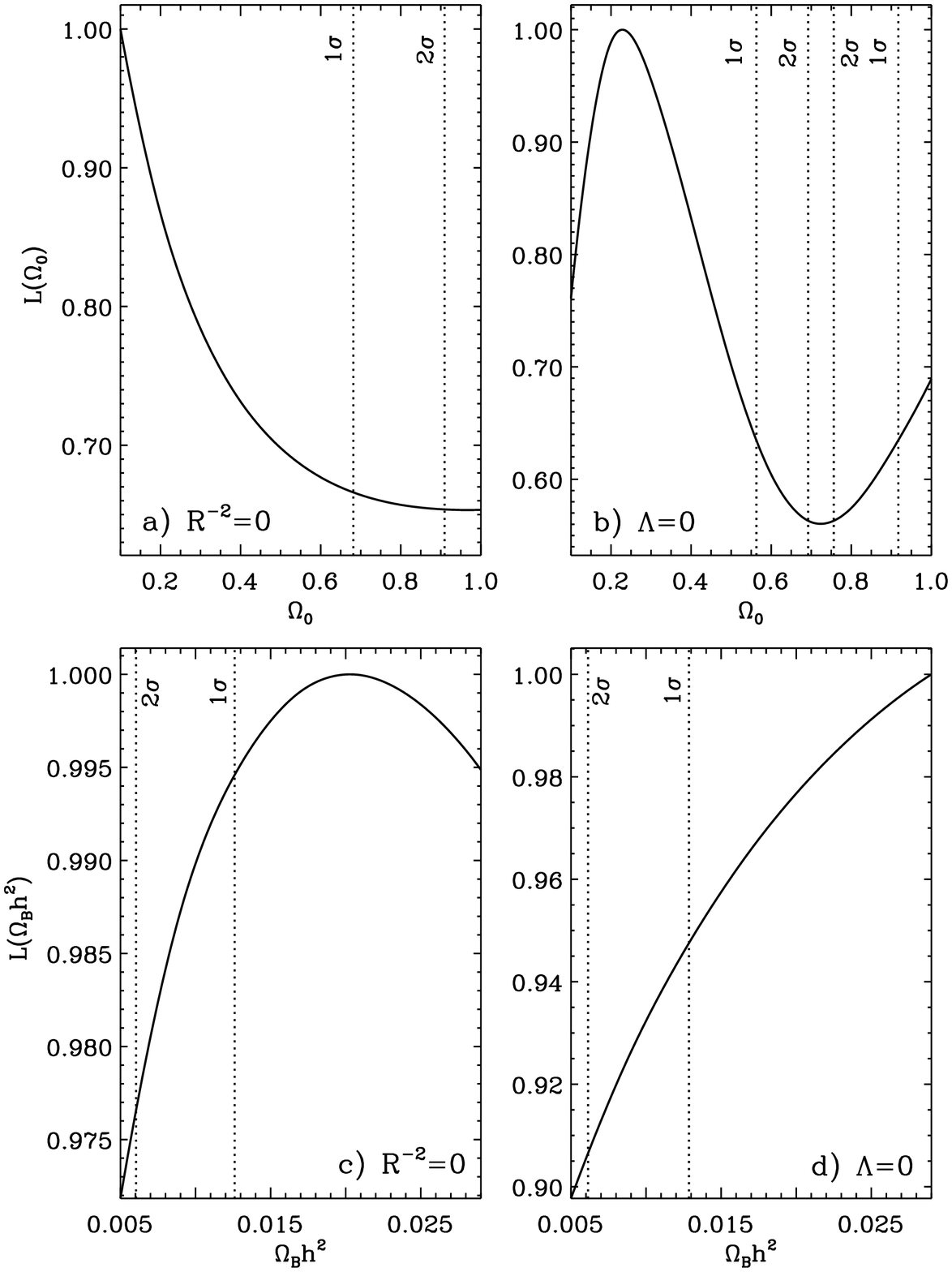}}
\mbox{\large Figure 6.1}
\end{figure}
\clearpage
\begin{figure}[t]
\centerline{\epsfysize=8.0truein \epsffile{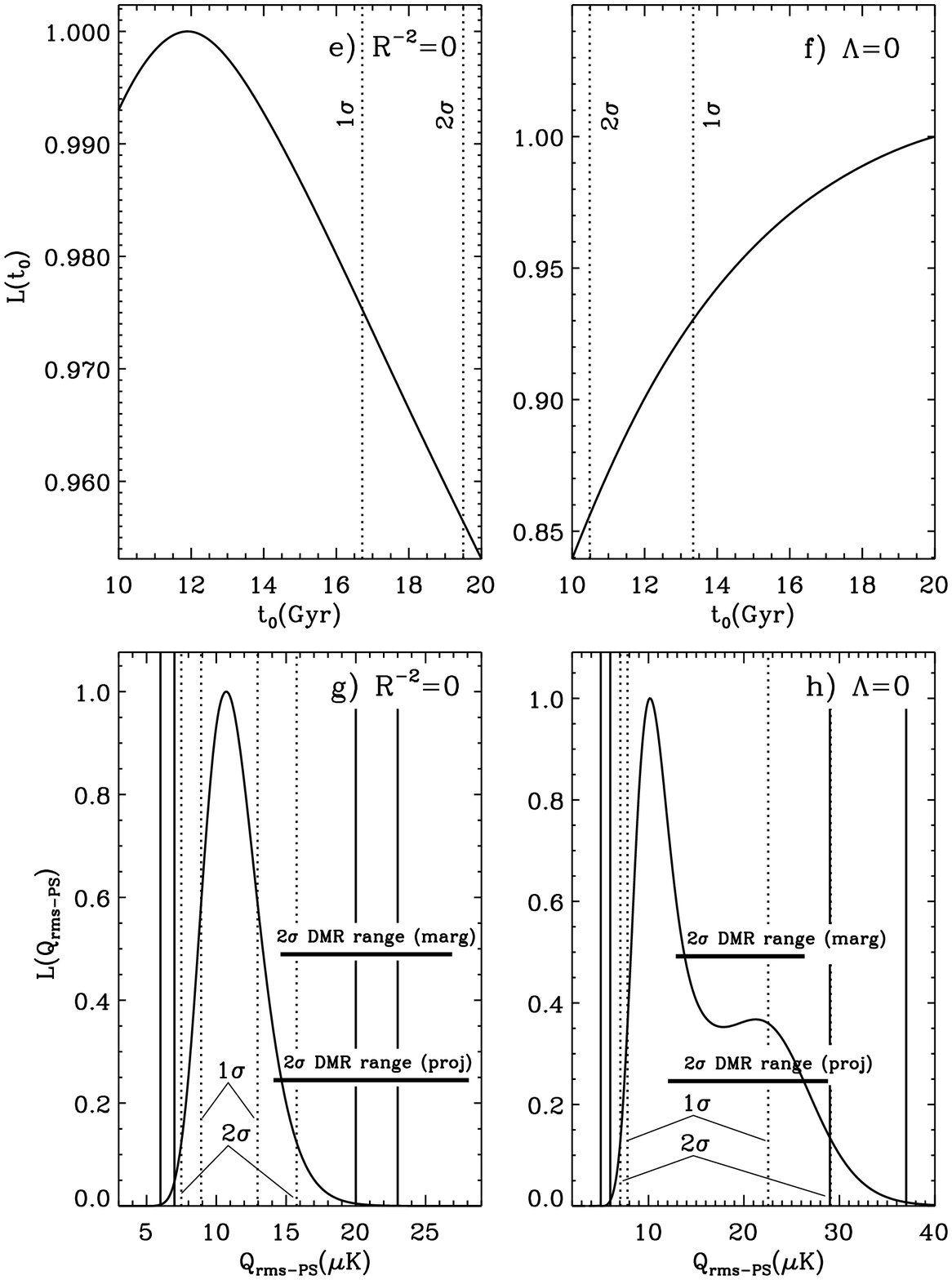}}
\mbox{\large Figure 6.2}
\end{figure}
\clearpage


\clearpage
\begin{references}
  \reference{Bartelmann} Bartelmann, M., Huss, A., Colberg, J.M., 
  Jenkins, A., \&\ Pearce, F.R. 1998, A{\&}A, 330, 1
  \reference{Bond} Bond, J.R., Efstathiou, G., Lubin, P.M., \&\
  Meinhold, P.R. 1991, \prl, 66, 2179
  \reference{BGT95} Bucher, M., Goldhaber, A.S., \&\ Turok,
  N. 1995, Phys. Rev. D, 52, 3314
  \reference{Bunn94} Bunn, E., White, M., Srednicki, M., \& Scott, D. 1994,
  \apj, 429, 1
  \reference{Burles} Burles, S., \&\ Tytler, D. 1998, ApJ, 499, 699
  \reference{Cavaliere} Cavaliere, A., Menci, N., \&\ Tozzi, P. 1998,
  ApJ, 501, 493
  \reference{Church97} Church, S.E., Ganga, K.M., Ade, P.A.R., Holzapfel, W.L.,
  Mauskopf, P.D., Wilbanks, T.M., \&\  Lange, A.E. 1997, \apj, 484, 523
  \reference{Cole} Cole, S., Hatton, S., Weinberg, D.H., \&\ Frenk, C.S.
  1998, MNRAS, in press
  \reference{Cole97} Cole, S., Weinberg, D.H., Frenk, C.S., \& Ratra, B.
  1997, \mnras, 289, 37
  \reference{Coles} Coles, P., Pearson, R.C., Borgani, S., Plionis, M., 
  \&\ Moscardini, L. 1998, MNRAS, 294, 245
  \reference{deB93} de Bernardis, P., et al. 1993, A{\&}A, 271, 683
  \reference{deB94a} de Bernardis, P., et al. 1994a, ApJ, 422, L33 (deB94)
  \reference{deB94b} de Bernardis, P., de Gasperis, G., Masi, S., \&\
  Vittorio, N. 1994b, ApJ, 433, L1
  \reference{Eke} Eke, V.R., Cole, S., Frenk, C.S., \&\ Henry, J.P. 1998,
  MNRAS, in press
  \reference{Fukugita} Fukugita, M., Hogan, C.J., \&\ Peebles, P.J.E. 1998,
  ApJ, in press
  \reference{GRGS} Ganga, K., Ratra, B., Gundersen, J.O., \&\ 
  Sugiyama, N.~1997a, \apj, 484, 7 (GRGS)
  \reference{Ganga97b} Ganga, K., Ratra, B., Church, S.E., Sugiyama, N.,
  Ade, P.A.R., Holzapfel, W.L., Mauskopf, P.D., \&\ Lange, A.E. 1997b, ApJ, 
  484, 517
  \reference{Ganga98} Ganga, K., Ratra, B., Lim, M.A., Sugiyama, N., \&\
  Tanaka, S.T. 1998, ApJS, 114, 165
  \reference{Gorski98} G\'orski, K.M., Ratra, B., Stompor, R., Sugiyama, N.,
  \&\ Banday, A.J. 1998, ApJS, 114, 1
  \reference{Gott} Gott, J.R. 1982, Nature, 295, 304
  \reference{Gratton} Gratton, R.G., Fusi Pecci, F., Carretta, E., 
  Clementini, G., Corsi, C.E., \&\ Lattanzi, M. 1997, ApJ, 491, 749
  \reference{Gundersen} Gundersen, J.O.,~\etal~1995, \apj, 443, L57
  \reference{Guth} Guth, A. 1981, Phys. Rev. D, 23, 347
  \reference{GW} Guth, A.H., \&\ Weinberg, E.J. 1983, Nucl. Phys. B, 212, 321
  \reference{Harrison} Harrison, E.R. 1970, Phys. Rev. D, 1, 2726
  \reference{Jenkins} Jenkins, A., et al. 1998, ApJ, 499, 20
  \reference{Kazanas} Kazanas, D. 1980, ApJ, 241, L59
  \reference{Lim96} Lim, M.A., et al. 1996, ApJ, 469, L69
  \reference{Masi95} Masi, S., et al. 1995, ApJ, 452, 253
  \reference{Masi96} Masi, S., de Bernardis, P., De Petris, M., Gervasi, M., 
  Boscaleri, A., Aquilini, E., Martinis, L., \&\ Scaramuzzi, F. 1996, ApJ,
  463, L47
  \reference{Merchan} Merch\'an, M.E., Abadi, M.G., Lambas, D.G., \&\ 
  Valotto, C. 1998, ApJ, 497, 32
  \reference{Netterfield} Netterfield, C.B., Devlin, M.J., Jarosik, N., 
  Page, L., \& Wollack, E.J. 1997, ApJ, 474, 47
  \reference{Palumbo} Palumbo, P., Aquilini, E., Cardoni, P., de Bernardis, P., 
  De Ninno, A., Martinis, L., Masi, S., \&\ Scaramuzzi, F. 1994, Cryogenics, 
  34, 1001
  \reference{Park} Park, C., Colley, W.N., Gott, J.R., Ratra, B., 
  Spergel, D.N., \&\ Sugiyama, N. 1998, ApJ, 506, in press
  \reference{Peebles70} Peebles, P.J.E., \&\ Yu, J.T. 1970, \apj, 162, 815
  \reference{Platt} Platt, S.R., Kovac, J., Dragovan, M., Peterson, J.B., 
  \& Ruhl, J.E.  1997, \apj, 475, L1
  \reference{Ratra98} Ratra, B., Ganga, K., Sugiyama, N., Tucker, G.S., 
  Griffin, G.S., Nguy{\^e}n, H.T., \&\ Peterson, J.B. 1998, ApJ, 505, in press 
  \reference{Ratra94} Ratra, B., \&\ Peebles, P.J.E. 1994, \apj, 432, L5
  \reference{Ratra95} Ratra, B., \&\ Peebles, P.J.E. 1995, Phys. Rev. D, 
  52, 1837
  \reference{Ratra97} Ratra, B., Sugiyama, N., Banday, A.J., \&\ 
  G\'orski, K.M. 1997, \apj, 481, 22
  \reference{Reid} Reid, I.N. 1997, AJ, 114, 161
  \reference{Sato1981a} Sato, K. 1981a, Phys. Lett. B, 99, 66
  \reference{Sato1981b} Sato, K. 1981b, MNRAS, 195, 467
  \reference{Somerville} Somerville, R.S., \&\ Primack, J.R. 1998, MNRAS,
  submitted
  \reference{Stompor94} Stompor, R. 1994, A{\&}A, 287, 693
  \reference{Stompor97} Stompor, R. 1997, in Microwave Background
  Anisotropies, ed. F.R. Bouchet, R. Gispert, B. Guiderdoni, \& J. Tran 
  Thanh Van (Gif-sur-Yvette: Editions Frontieres), 91
  \reference{Sugiyama95} Sugiyama, N. 1995, ApJS, 100, 281
  \reference{Tanaka96} Tanaka, S.T.,~\etal~1996, \apj, 468, L81
  \reference{Tucker93} Tucker, G.S., Griffin, G.S., Nguy\^en, H.T., 
  \&\ Peterson, J.B. 1993, \apj, 419, L45 
  \reference{Yamamoto} Yamamoto, K., Sasaki, M., \&\ Tanaka, T. 1995,
  \apj, 455, 412
  \reference{Zeldovich} Zel'dovich, Ya.B. 1972, MNRAS, 160, 1P
\end{references}
\end{document}